\documentclass[10pt, conference]{IEEEtran}
\usepackage{amsmath,amssymb,subfig,color,graphicx, proof,colortbl,mathrsfs,mathbbol,multirow,styles/dsfont,soul,float,balance}
\definecolor{Gray}{gray}{0.88}

\newtheorem{lemma}{Lemma}

\newtheorem{corollary}{Corollary}
\newtheorem{remark}{Remark}
\newtheorem{example}{Example}

\newcommand{\depths}{\mathcal{D}}

\newcommand{\levels}{\mathcal{L}}

\newcommand{\vulnerabilities}{\mathcal{V}}

\newcommand{\targets}{\mathcal{T}}
\newcommand{\DS}{{\Delta S}}
\newcommand{\DC}{{\Delta C}}

\newcommand{\1}{\langle}
\newcommand{\2}{\rangle}

\newcommand{\pD}{Defender~}
\newcommand{\pDw}{Defender}
\newcommand{\Defender}{D}
\newcommand{\DefMixed}{\Phi}
\newcommand{\pA}{Attacker~}
\newcommand{\pAw}{Attacker}
\newcommand{\Attacker}{A}
\newcommand{\AttackerMixed}{\Theta}
\newcommand{\DefMat}{\mathds{A}}
\newcommand{\NegDefMat}{\mathds{-A}}
\newcommand{\AttackerMat}{\mathds{E}}

\newcommand{\boxsize}{1in}

\begin{document}
\title{Comparing Decision Support Approaches \\ for Cyber Security Investment}
\author{

\IEEEauthorblockN{
Andrew Fielder\IEEEauthorrefmark{1}, 
Emmanouil Panaousis\IEEEauthorrefmark{2}, 
Pasquale Malacaria\IEEEauthorrefmark{3},\\ 
Chris Hankin\IEEEauthorrefmark{1}, 
and
Fabrizio Smeraldi\IEEEauthorrefmark{3}}

\IEEEauthorblockA{
\IEEEauthorrefmark{1}
Imperial College London, UK}

\IEEEauthorblockA{
\IEEEauthorrefmark{2}
University of Brighton, UK}

\IEEEauthorblockA{
\IEEEauthorrefmark{3}
Queen Mary University of London, UK}
}

\maketitle
\thispagestyle{plain}
\pagestyle{plain}
\begin{abstract}
When investing in cyber security resources, information security managers have to follow effective decision-making strategies. 
We refer to this as the cyber security investment challenge. 
In this paper, we consider three possible decision-support methodologies for security managers to tackle this challenge. 
We consider methods based on game theory, combinatorial optimisation and a hybrid of the two. 
Our modelling starts by building a framework where we can investigate the effectiveness of a cyber security control regarding the protection of different assets seen as targets in presence of commodity threats. 
In terms of game theory we consider a 2-person control game between the security manager who
has to choose among different implementation levels of a cyber security control, and a commodity attacker who chooses among different targets to attack.  
The pure game theoretical methodology consists of a large game including all controls and all threats.
In the hybrid methodology the game solutions of individual control-games along with
their direct costs (e.g. financial) are combined with a knapsack algorithm to derive an optimal investment strategy. 
The combinatorial optimisation technique consists of a multi-objective multiple choice knapsack
based strategy.
We compare these approaches on a case study that was built on SANS top critical controls. 
The main achievements of this work is to highlight the weaknesses and strengths of different investment methodologies for cyber security, the benefit of their interaction, and the impact that indirect costs have on cyber security investment.
\end{abstract}
\section{Introduction}
One of the biggest issues facing organisations today is how they are able to defend themselves from potential cyber attacks.  The range and scope of these unknown attacks create the need for organisations to prioritise the manner in which they defend themselves.  
With this each organisation needs to consider the threats that they are most at risk from and act in such a way so as to reduce the vulnerability across as many relevant vulnerabilities as possible.  
This is a particularly difficult task that many Chief Information Security Officers (CISOs) are not confident in achieving, with  only 24\% of CISOs considering themselves very confident of preventing attacks according to Deloitte and NASICO \cite{Deloitte}.  
In this report 86\% of CISOs were concerned that the biggest issue facing their ability to successfully defend their systems was down to a ``lack of sufficient funding''.

It is this perceived lack of sufficient funding that this work wishes to address.  
From our work with Small-Medium Enterprises (SMEs), we have identified that they are heavily restricted with the available funding for cyber security, generally working with a fixed budget with little to no additional funding being made available for cyber security purposes.  
It is generally perceived that this budget is insufficient for them to cover all of the vulnerabilities that their system may have.  
In this way organisations have to make trade-offs with regard to how they defend their systems.

When an organisation is making the decisions regarding the defence of their network, they generally have to consider two critical factors, the cost of implementing a particular defence and the impact that defence has on the business.  The first of these has been discussed, stating that a company can only implement defences that are within their limited budget, considered the \emph{Direct Cost} of the defence.  
However we question whether the apparently most optimal defence based solely on direct costs is the correct choice for an organisation. 
The reason behind this lies with the second criteria, such that the manner in which a defence is implemented will likely have some effect on either the operation of the system or the users of the system.  
These effects may cause a reduction in the speed that tasks can be performed by users or by a weakening of the defence caused by users circumventing the controls in order to more easily perform their required tasks.  
We consider that these factors create additional \emph{indirect costs} for implementing a given defence.
These two factors are at the core or our work into the decision support of \emph{how to use the limited financial budget available to best protect against cyber attacks}.

\subsection{Contributions}
This work proposes a two stage model designed to aid security managers with decisions regarding the optimal allocation of a cyber security budget. 
And it provides a detailed analysis of the model that was first proposed by Panaousis et al. \cite{Manos2014}.

We analyse the two stages of the model by first presenting an overview of the environment from which we define the problem of \emph{cyber security investment}, identifying a unique manner for reasoning about the \emph{targets} that a potential attacker has, and the defences associated with those targets.  
This is done by considering the physical location of a data asset, which needs to be protected, as well as the degree to which a particular defence, herein referred to as a \emph{control}, is implemented. 

We use the above environment to formulate control-games, which analyse how well each given control performs at different \emph{degrees of implementation} (i.e.\,levels).  
We compute the Nash Equilibrium condition in control-games, and we motivate the trade-offs required with the indirect costs.  
The Nash Equilibrium of a control-game dictates the most efficient manner, in which, a control should be implemented.

The solution to each control-game alone is insufficient in dictating the optimal allocation of an organisation's cyber security budget.  
So to identify the best way to allocate a budget, we formalise the problem as a \emph{multi-objective multiple choice Knapsack problem}, as proposed in \cite{Smeraldi:ACySe:2014}.  

We motivate the use of this methodology by comparing the two-stage model to two alternative methods.  
Firstly, we model the scenario as an one shot game that aims to optimise the defense including \emph{direct costs}, and secondly a Knapsack problem that considers only pure strategies for each control level including indirect costs.  
In both cases we highlight where our proposed method is able to outperform alternative methods.

\subsection{Outline}

Section \ref{sec_environment} introduces \emph{targets} i.e.\,vulnerability and the associated potential loss, \emph{controls}, effectiveness of controls on the potential loss as well as indirect costs associated to implement controls at different levels. 
In Section \ref{sec_game} those notions are used to build a game model, and sections \ref{sec_pure_strategies} to \ref{sec_non_zero} develop a basic analysis of these games.  
Sections \ref{sec_toy_game} to \ref{sec_numeric_illustration} present a toy 2x2 game example with a single control with two implementation levels and two targets. 
This aims to provide a feel for these games and what elements determine the equilibria.

In Section \ref{sec:investment} we discuss general games where the defender has available a single control and the attacker can choose a set of vulnerabilities: we provide an interpretation of mixed strategies for these games and explain how these game solutions can be used with a Knapsack algorithm.
Section \ref{sec_full_game} illustrates why a single game comprising all possible controls doesn't provide good security guarantees and so is not a suitable investment methodology. Section \ref{sec_pure_knap} introduces the particular Knapsack we think is relevant for this work, that is 0-1 Multiple Choice, Multi-Objective Knapsack. The items relevant to this Knapsack are game solutions: this constitutes the hybrid methodology presented in section \ref{sec_hybrid}.

In Section \ref{sec_case} we develop a case study based on the SANS top critical controls. We illustrate a mapping from the SANS controls' descriptions into our framework and based on this mapping we run a simulation to calculate the security effectiveness of the two Knapsack methodologies. The simulation shows that the hybrid solution is more flexible and provides higher security guarantee.
\section{Related Work}
The work of Anderson \cite{Anderson:ACSAC:2001} considers the traps that defenders may fall into in finding bugs and protecting their systems, where it only needs to be a single unseen vulnerability that exposes the whole of a network.  
The approach taken in this work is to model attackers using \emph{commodity} attacks against SMEs, where the attacker is using commonly available attack vectors against known defendable vulnerabilities.  
While this doesn't negate the possibility of zero-day vulnerabilities, it removes the expectation that it is in the best interest of either player to invest heavily in order to either find a new vulnerability or be able to protect against these unknown vulnerabilities. 
In his work, Anderson considers that the management of information security is a more difficult problem than initially considered as there are often deeper issues, such as politics, that need to be addressed.  

Important to the modelling is the concept that the defenders have to attempt to \emph{defend everywhere}.  This is due to the fact that attackers can strike anywhere they wish. 
We can highlight this observation by assuming that the defence provided by an optimal budget allocations can only be considered as strong as the defence of the \emph{weakest target}. This is because the weakest target is at most risk from an attacker who can potentially attack anywhere.
Our approach is quite different to Anderson's as we focus on developing \emph{cyber security decision support tools} to assist security managers on how to spend a cyber security budget in terms of different controls acquisition and implementation.

Our work has been partially influenced by a recent contribution within the field of physical security \cite{Kiekintveld:AAMAS:2013}, where the authors address the problem of finding an optimal defensive coverage.
The latter is defined as the one maximising the worst-case payoff over the targets in the potential attack set.
One of the main ideas of this work we adopt here is that \emph{the more we defend the less rewards the attacker receives}. 

Alpcan \cite{Alpcan:Book:2012} (p.\,134)\,discusses the importance of studying the quantitative aspects of risk assessment with regard to cyber security in order to better inform decisions makers.  
This kind of approach is taken in this work where we provide an analytical method for deciding the level of risk introduced by different vulnerabilities, and the impact that different security controls have in mitigating these risks. 
By studying the incentives for risk management, Alpcan \cite{Alpcan:NTMS:2012} develops a game-theoretic approach that optimises the investment in security across different autonomous divisions of an organisation, where each of the divisions is seen as a greedy entity.
Furthermore, Alpcan et al. examine in \cite{Alpcan:CRiSIS:2009} security risk dependencies in organisations, and they propose a framework which ranks the risks by considering the different complex interactions. 
This rank is dictated by an equilibrium that is derived by a Risk-Rank algorithm.

Saad et al. \cite{Saad:ICIMP:2010} model cooperation among autonomous parts of an organisation that have dependent security assets, and vulnerabilities for reducing overall security risks, as a cooperative game.
In \cite{Bommannavar:ICC:2011} Bommannavar et al. capture risk management in a quantitative framework which aids decision makers upon allocation of security resources. 
The authors use a dynamic zero-sum game to model the interactions between attacking and defending players.
A Markov model, in which states represent probabilistic risk regions and transitions, has been defined. 
The authors use Q-learning to cope with scenarios when players are not aware of the different Markov model parameters.

Fielder et. al.\,\cite{Fielder:IFIPSEC:2014} investigate \emph{how to optimally allocate the time for security tasks for system administrators}. 
This work identifies how to allocate the limited amount of time, which a system administrator has, to work on the different security related tasks for an organisation's data assets.

One of the initial works studying the way to model investment in cyber security is published by Gordon and Loeb \cite{Gordon:TISSEC:2002}.  
The authors consider the optimum level of investment given different levels of information security level.  
The authors propose a model in which for any given vulnerability there are different levels of information security that can be implemented, where a higher level of information security will cause the expected loss to that particular vulnerability to drop. 
This is modelled as a function of the security level's responsiveness to an increasing vulnerability in reducing loss.  
In our model, here, we consider a single value for a  vulnerability, and then for each control there are a number of levels of implementation, which represent the information security levels proposed by Gordon and Loeb.
The main message of this work is that to maximise the expected benefit from information security investment, an organisation should spend only a small fraction of the expected loss due to a security breach.

The work published in \cite{Johnson:Springer:2012} examines the weakest target game which refers to the case where an attacker is always able to compromise the system target with the lowest level of defence, and not to cause any damage to the rest of the targets. 
The game-theoretic analysis, which the authors have undertaken, shows that the game leads to a conflict between pure economic interests and common social norms. 
While the former are concerned with the minimisation of cost for security investments, the latter imply that higher security levels are preferable.
Cavusoglu et. al. \cite{Cavusoglu:JMIS:2014}\,compare a decision theory based approach to game-theoretic approaches for investment in cyber security. 
Their work compares a decision theory model to both simultaneous and sequential games.  
The results show that the expected payoff from a sequential game is better than that of the decision theoretic method, however, a simultaneous game is not always better.

Recent work on cyber security spending has been published by Smeraldi and Malacaria \cite{Smeraldi:ACySe:2014}.
The authors identify the optimum manner in which investments can be made in a cyber security scenario given that the budget allocation problem is most fittingly represented as a multi-objective Knapsack problem.
Cremonini and Nizovtsev, in \cite{Cremonini:WorkingPaper:2006}, have developed an analytical model of the attacker's behaviour by using cost-benefit analysis, and therefore considering rewards and costs of achieving different actions. 
One issue that we factor into this work is that \emph{security comes at a cost that is greater than that of the price of implementing a policy}.  
Al-Humaigani and Dunn proposed a model of Return on Security Investment (ROSI) \cite{Dunn2012}, where the authors define the return on investment of an attack as a function of eleven factors, which comprise of \emph{direct costs} for implementing a security tool, \emph{indirect costs} of having that \emph{security tool} in place, as well as the cost to the company should there be a breach (i.e.\,\emph{damage}).
Wang et al. note that game-theoretic approaches to cyber security suffer from the fact that ``the rationality of hackers is hard to be captured by a model, because they may be motivated by different value systems'' \cite{Wang2008}.  
While the authors do not argue on the rationality of the attacker, but the idea that imposing on them a similar set of values as a defender is not adequate. 
Previous work we have conducted in this area notes that the reward for the attacker is in line with the loss of the defender by the way of an affine transformation \cite{Fielder:IFIPSEC:2014}.
This was done to represent the loss of value that an attacker gets from the data that has been stolen, when compared to the value to the defender.

Demetz and Bachlechner \cite{Demetz:WEIS:2012} provide a survey of models that have been proposed for the study of economic viability of tools for security policy and configuration.  
The authors identify a series of requirements that a security investment tool should contain.  
We compare our approach to the conditions set out by Denetz and Bachlechner's:

\begin{itemize}
\item Financial Measures - The optimisation method looks to take into consideration the financial constraints of the organisation and identifies what should be purchased given the range of possible budgets.  
An organisation is then able to select an appropriate set of controls given their financial constraints and threat tolerance levels. 

\item Non-Financial Measures - One of the key features of a multi-level model of a control implementation is that, it is possible to clearly identify non-financial measures such as System Performance or Staff Morale that will be impacted as a result of implementing certain controls or extending the reach that some controls have (such as surveillance).

\item Support One-Time Costs and Benefits - One of the direct costs is Capital Costs, where the capital costs for an organisation incurred for implementing the control.

\item Support Running Costs and Benefits - As with the One-Time costs, the model supports the inclusion of running costs into these direct costs as well, primarily as labour costs.  Additionally any non-financial costs incurred in the actual implementation are represented in the indirect costs.

\item Does not explicitly consider Attacks - The game-theoretic model that is presented here is defined in such a way that it allows for the representation of a control that is capable of mitigating any number of attacks.  
However, this goal focuses on policy and configuration that will not only protect against attacks, but will also work with security breaches that are not related to the attacks.

\item Consider the Network Effects of Investments - Within the scope of the model considered, there is no direct consideration of the additional benefits to other organisations from the implementation of a given security policy given by our model. 
\end{itemize}

While the work we provide covers many of the aspects designated to be an effective security tool, it notably lacks the aspects that relate to non-attack related issues of security such as unintentional loss or network benefits.
\section{Model Definition}
In this section we use game theory to model the interactions between two players; \pD and \pAw.
\pD is the cyber security manager in an SME, and her overall objective is to defend the organisation's assets from cyber theft, mitigate any 
potential business disruption, and maintain the organisation's reputation.

On the other hand, \pA is a cyber hacker who tries to subvert the system to her own end, by launching commodity cyber-attacks against the organisation \pD is working for. 
Commodity cyber-attacks are based on capabilities and techniques that are  available on the internet, where the attack tool can be purchased therefore the adversaries do not develop the attack themselves, and they can only configure the tool for their own use.

\begin{figure*}
\centering
  \includegraphics[width=5.5in]{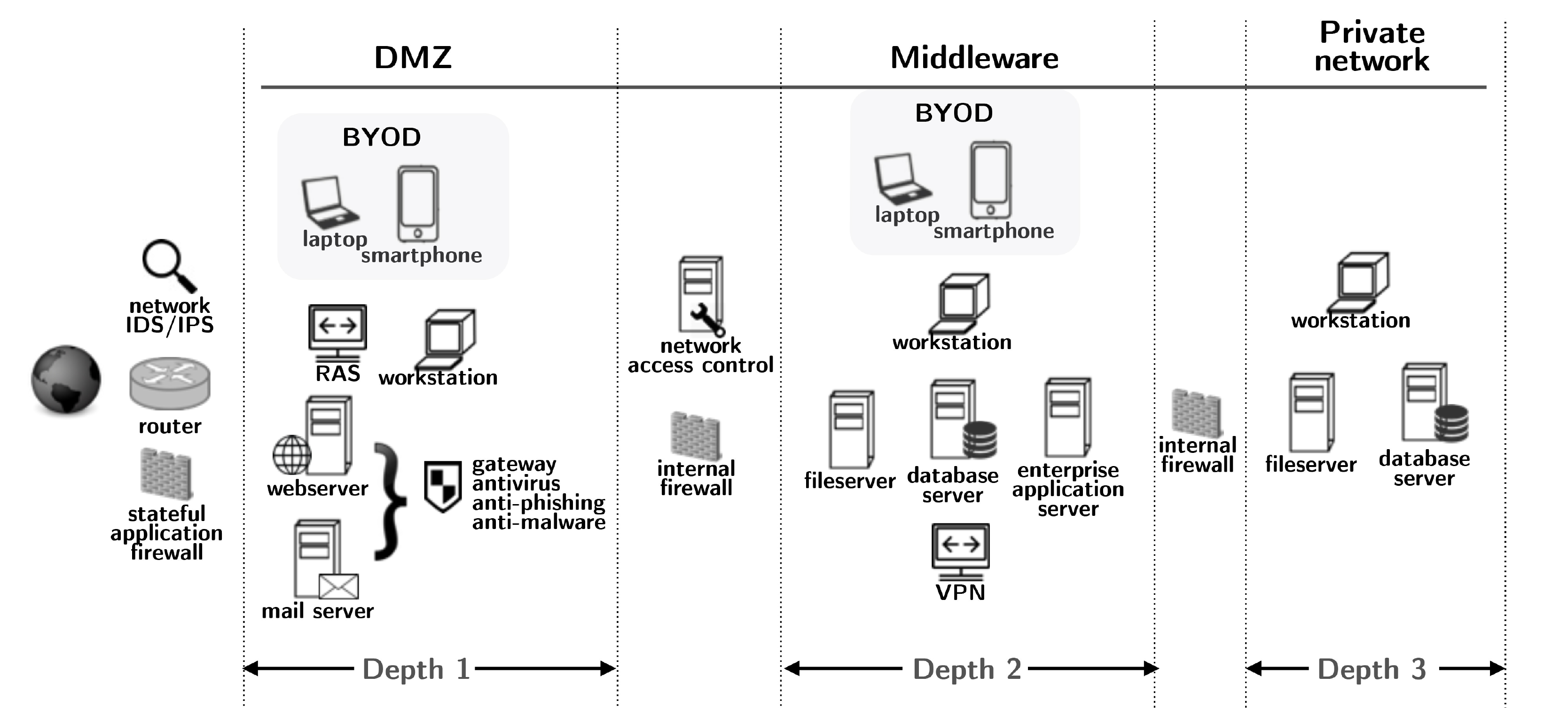}
  \caption{Illustration of our environment \cite{Manos2014}.}\label{fig:environment}
\end{figure*}

First, we present the cyber security model and environment of the game, and introduce the two players. 
The model of environment and game presented here were initially proposed in \cite{Manos2014}. 
We then describe the sets of their pure strategies and the payoffs associated with each of them.
We also discuss how mixed strategies are represented in this game, and we provide the expected payoffs of \pD and \pA given any pair of mixed strategies.
Fig.\,\ref{fig:environment} illustrates our environment.

\subsection{Environment}
\label{sec_environment}
In our model, \pD work as a cyber security manager in an SME with an available cyber security budget $B$, and she wants to invest in implementing cyber security controls to protect the organisation's data assets against \emph{commodity attacks}.  

Each control can be implemented at a different level.
Note that the higher the level the greater the degree to which the control is implemented.
After its implementation, each control brings some security benefits to the system, but it is also associated with indirect and direct costs. 
The challenge \pD has to address is how to decide upon implementation of the different cyber security controls against \emph{commodity attacks}, given a limited budget $B$, and other preferences the organisation has in terms of risks and indirect costs.
In the following we discuss the different components of the model, and we define appropriate terminology and notations, which are consistent throughout this article.

\subsubsection{Asset Depth}
We define the \emph{depth} of a data asset as the location of this asset within the organisation's structure following the rule: \emph{the higher the depth is, the more confidential data the asset holds}. 
In other words, a depth determines the importance of the data asset that the organisation loses if a commodity attack (herein referred to as attack) is successful.
In this paper, we specify that data assets that located at the same are depth, worth the same value to \pDw's firm.

\subsubsection{Cyber Security Targets}
We denote the set of cyber security targets within an organisation by $\targets := \{t_i\}$, the set of vulnerabilities threatened by commodity attacks by $\vulnerabilities := \{v_z\}$, and the set of depths by $\depths := \{d_x\}$.
A \emph{cyber security target} is defined as a $(vulnerability, depth)$ pair; formally 
\begin{equation}\label{eq_target}
	t_i := (v_z, d_x).
\end{equation}
And it abstracts any data asset, located at $d_x$, that an attack threatens to compromise by exploiting $v_z$.
We specify that data assets located at the same depth and having the same vulnerabilities are abstracted by the same target.

Each target is associated with an impact value which expresses the level of damage incurred to \pDw's organisation when \pA succeeds in their attack against that target.
The different impact factors can be \emph{data loss}, \emph{business disruption}, and \emph{reputation damage}.
Each impact factor depends on the depth $d_x$ that the attack targets.

Furthermore, there is a \emph{threat value} for each target. 
This can account, for instance, for the frequency of attacks launched against that target. 
Each software weakness (we use the terms weakness and vulnerability interchangeably) has some factors that can determine an overall score.


Let $I\colon\,\targets\rightarrow\mathbb{Z}^+$ be the random variable which takes targets $t_i$ to the impact value that the compromise of $t_i$ will have to the organisation, and let $T\colon\,\targets\rightarrow \mathbb{Z}^+$, be the random variable which takes target $t_i \in \targets$ to its threat value.
Given Definition\,(\ref{eq_target}) note that $I(t_i)$ depends on the depth $d_x$, and $T(t_i)$ depends on the vulnerability $v_z$. 

\subsubsection{Cyber Security Controls}
A \emph{cyber security control} is the defensive mechanism that \pD can be put in place to alleviate the risk from one or more attacks by reducing the probability of these attacks successfully exploiting vulnerabilities. 

\pD chooses to implement a control at a certain level for their organisation. 
We define the set of implementation levels of a control as $\levels := \{l_j\}$.
The higher the level the greater the degree to which the control is implemented.

Note that we abuse notation by setting $l_j = l, t_i = t, v_z = v$, and $d_x = d$.

An implementation level $l$ has a \emph{degree of vulnerability mitigation} on each target.
This is determined by the efficacy, in terms of cyber defence, of $l$ on this target.
For a pair $(l, t)$, which represents the level of implementation of a particular control, we define the random variable $E\colon\levels \times \targets \rightarrow [0,1)$, which takes a pair of $(l_,t)$ to the efficacy value of $l$ on $t$.
Here, we have postulated that $E(l,t) \neq 1$ due to the existence of 0-day vulnerabilities that \pA has the potential to exploit. 
Assume \pD implements a control at $l$ that has efficacy $E(l,t)$ on $t$.

We define the \emph{cyber security loss} random variable 
\[
S(l,t) = I(t) \, T(t) \, [1-E(l,t)].
\]
This is the expected damage (e.g. losing some data asset) that \pD suffers when $t$ is attacked and a control has implemented at level $l$.

While the implementation of a cyber security control strengthens the defence of \pDw's organisation, it is associated with two types of costs namely; \emph{indirect} and \emph{direct}.
Examples of indirect cost are System Performance Cost, Morale Cost, and Re-Training Cost.

%
%
%
For a level $l$ we express its indirect cost by the random variable $C\colon\levels \rightarrow \mathbb{Z^+}$.
From the above we can derive the overall loss of \pDw's organisation.
This is equivalent to the sum of the security damages inflicted by 
\pA and the indirect cost for implementing a control at a certain 
level. 
Formally, when \pD implements a control at some level $i$ then the 
expected loss of their organisation is derived by
\begin{equation*}\label{eq:alice_total_loss}
 \sum_{t} S(l,t) - C(l).
\end{equation*}
The implementation of a control, at some level, has a direct cost 
which refers to the budget the organisation must spend to this 
implementation.
For instance, we can split such direct cost into two categories, the 
Capital Cost and Labour Cost.
We express the direct cost of an implementation level $l$ by the 
random variable $\Gamma\colon\levels \rightarrow \mathbb{Z^+}$ that 
takes implementation levels to the monetary cost of the control 
implementation.
 
\begin{table*}[t]
\centering
\renewcommand{\arraystretch}{1.5}
\caption{Notation.\label{tab:notation}}
{
  \begin{tabular}{|clcl|}
  \hline
  $\mathcal{T}$ & set of cyber security targets &
  $\levels$ & set of implementation levels \\
 
  \rowcolor{Gray}
  $\mathcal{V}$ &  set of vulnerabilities &
  $\mathcal{D}$ & set of depths \\
  
  $T(t)$ & threat value of target $t$ &
  $I(t) $ & \parbox[c]{1.5in}{\vspace{0.05cm}
  impact of a successful attack against target $t$} \\
  
  \rowcolor{Gray}
  $S(l,t) $ & \parbox[c]{1.5in}{\vspace{0.05cm}
  security loss when target $t$ is compromised} &
  $E(l,t)$ & effectiveness of $l$ on $t$  \\
  
  $\theta_j$ & probability $t_j$ to be attacked &
  $\phi_j$ & probability $l_j$ to be selected \\
  
   \rowcolor{Gray}
   $C(l)$ & \parbox[c]{1.5in}{\vspace{0.05cm}indirect cost of $l$} &
   $\Gamma(l)$ & total direct cost of $l$  \\   
  \hline
\end{tabular}}
\end{table*}

\subsection{Game Characterization}
\label{sec_game}
In this article we formulate a two-player non-cooperative static game.
The players in our game are \pD (she represents any cyber security decision-maker) and \pA (she represents any cyber hacker who uses commodity attacks).
\pD defends their organisation's data assets by minimising expected cyber security losses with respect to the indirect costs, while the attacker \pA aims at benefiting from compromising \pDw's organisation data assets.

The \pD is choosing how to implement a cyber security control (i.e. at which level) and  \pA decides which target to attack to exploit its vulnerability at a certain depth.
Since we consider a simultaneous game \pA does not know the control implementation strategy and \pD does not know the attack strategy.
We refer to our games as \emph{control games} because the basis of our formulation is that \pD has one control at her disposal. 

In this article we formulate a zero-sum game. 
This represents scenarios where \pA aims at causing the maximum possible damage to \pDw.
We believe that if we consider a non-zero sum game then a specific threat model must be defined as well. 
Such a model could consider, for instance, some cost for \pA when undertaking an attack. 
However cost in terms of cyber attacks is tightly coupled with the profile of the attacker.
A consideration of a specific threat model would also have some influence on the way \pA sees the different targets as she is after specific goals based on her motivation (i.e, cyber crime, hacktivism, cyber espionage).
In this case, different \pA profiles could have been investigated.
In our work here, we have not investigated such profiles and our work is limited to a generic assumption that \pA is taking advantage of commodity attacks that she can purchase from online sources. 
In other word, we have assumed a set of attack methods that \pA can choose from but we have not postulated anything about their motivations.

\subsection{Pure-Strategy Sets}
\label{sec_pure_strategies}
For a given cyber security control, \pD can choose to implement the control at level $l \in \levels$ and therefore her pure strategy set coincides with $\levels$.
The \pA selects a vulnerability to exploit at a certain depth.
Formally, \pA chooses $t = \1 v, d \2 \in \targets$.
Thus the pure strategy set of \pA coincides with $\targets$.

\subsection{Payoffs}
\label{sec_payoffs}
Given that the pure strategy sets of the players are $\levels$ and $\targets$ then \pD has $m$ pure strategies and \pA has $n$, correspondingly. 
We denote by $G := \1\DefMat, \AttackerMat\2$ an $m \times n$ bi-matrix cyber security game where \pD (i.e.\,row player) has a payoff matrix $\DefMat \in \mathbb{R}^{m \times n}$ and the payoff matrix of \pA (i.e.\,the column player) is denoted by $\AttackerMat \in \mathbb{R}^{m \times n}$.

\pD chooses as one of her pure strategies one of the rows of the payoff bi-matrix $\1\DefMat, \AttackerMat\2 := [ (a_{lt} ,e_{lt}) ]_{l,t}$.
For any pair of strategies $(l,t)$, \pD and \pA have payoff values equivalent to $a_{lt}$ and $e_{lt}$, given by
\begin{eqnarray}\label{eq:alice_payoff}
	&& a_{lt} := S(l,t) - C(l) \nonumber\\
	&& e_{lt} := -S(l,t) + C(l).\nonumber
\end{eqnarray}

Tables \ref{tab:defenders_payoff} and \ref{tab:attackers_payoff} are the player's payoff matrices.

\begin{table}[h]
  \renewcommand{\arraystretch}{1.3}
  \centering
  \caption{\pDw's payoff matrix.\label{tab:defenders_payoff}}{
  \begin{tabular}{|c|c|c|}
   \hline
   & $t$  &   $t'$ \\  \hline
   $l$ & $S(l,t) - C(l)$ & $S(l,t') - C(l)$  \\
   $l'$ & $S(l',t) - C(l')$ & $S(l',t') - C(l')$ \\
\hline
\end{tabular}}
\end{table}

\begin{table}[h]
\centering
  \renewcommand{\arraystretch}{1.3}
  \caption{\pAw's payoff matrix.\label{tab:attackers_payoff}}{
  \begin{tabular}{|c|c|c|}
   \hline
   & $t$  &   $t'$ \\  \hline
   $l$ & $-S(l,t) + C(l)$ & $-S(l,t') + C(l)$  \\
   $l'$ & $-S(l',t) + C(l')$ & $-S(l',t') + C(l')$ \\
\hline
\end{tabular}}
\end{table}

\subsection{Representation of Mixed Strategies}
\label{sec_mixed_strategies}
A player's mixed strategy is a distribution over the set of their pure strategies. 
The representation of \pDw's mixed strategy space is a finite probability distribution over the set of the different control implementation levels $\{l_1, \dots, l_{m}\}$.
For \pAw, the representation of their mixed strategy space is a probability distribution over the different targets $\{t_1, \dots, t_{n}\}$.

In this paper we are interested in how different control implementation levels are combined in a proportional manner to give an implementation plan for this control.
We call this a \emph{cyber security plan}. 
This allows us to examine advanced ways of mitigating vulnerabilities.
A cyber security plan is a probability distribution over different cyber security processes.
When investing in cyber security we will be looking into the direct cost of each cyber security plan which is derived as a combination of the different costs of the cyber security processes that comprise this plan.

\subsubsection{\pDw's Mixed Strategies}
We define \pDw's mixed strategy as the probability distribution $\DefMixed =[\phi_1, \dots, \phi_{m} ]$.
This expresses a cyber security plan, where $\phi_j$ is the probability of implementing the control at $l_j$.
A cyber security plan can be realised as advice to \pD on how to implement a cyber security control 
by combining different implementation levels.
Although this assumption complicates our analysis at the same time it allows us to reason about equilibria of the control game therefore providing a more effective strategy for \pDw.
We claim that our model is flexible thus allowing \pD to interpret mixed strategies in different ways to satisfy their requirements.

\subsubsection{\pAw's Mixed Strategies}
A mixed strategy of \pA is defined as a probability distribution over the set $\{v_1, \dots, v_\nu\} \times \{d_1, \dots, d_\Delta\}$.
In a simpler form, the mixed strategy of \pA is a probability distribution over the different targets and it is denoted by $\AttackerMixed = [\theta_{1}, \dots, \theta_{n}] $, where $\theta_i$ is the probability of the adversary attacking $t_i$. 

\subsubsection{Payoffs for Mixed Strategies}\label{payoffs}
When both players choose a pure strategy randomly according to the probability distributions determined by $\DefMixed$ and $\AttackerMixed$, the expected payoffs to \pD and \pA are
\begin{align*}
  J_{\Defender}(\DefMixed, \AttackerMixed) & := \sum_{i=1}^{n} \sum_{j=1}^{m} \phi_j \, a_{ij} \, \theta_i \nonumber\\
  J_{\Attacker}(\DefMixed, \AttackerMixed) & := \sum_{i=1}^{n} \sum_{j=1}^{m} \phi_j \, e_{ij} \, \theta_i. 
\end{align*}

\subsection{Best Responses Analysis}
\label{sec_br_analysis}
For the remainder of this section, we analyse a specific control game.  We assume that for a specific target $t$, \pD has only two possible levels at her disposal namely $l$, and $l'$ (e.g. performing penetration testing rarely during a year or often), to implement a control.
We define 
\begin{equation*}\label{deltaS}
	\DS(t) := S(l',t) - S(l,t)
\end{equation*} 
\begin{equation*}\label{deltaC}
	\DC := C(l') - C(l)
\end{equation*}
$\DS(t)$ is the reduction in damage when $l'$ is chosen, and $\DC$ is the extra indirect cost of $l'$ over $l$.

\begin{lemma}
When the reduction in damage achieved by $l'$ over $l$ is higher than the extra indirect cost that $l'$ introduces, \pD chooses $l'$.
\end{lemma}
\begin{IEEEproof}
If the reduction in damage achieved by $l'$ over $l$ is higher than 
the extra indirect cost that $l'$ then $\Delta S(t) > \Delta C$. This 
can be broken down as, $S(l',t) - S(l,t) > C(l') - C(l) \Leftrightarrow S(l',t) - C(l') > S(l,t) - C(l) \Leftrightarrow a_{l't} > a_{lt}$.  
Therefore, the \pD is 
incentivised to pick $l'$ as it has a higher utility.
\end{IEEEproof}



\begin{lemma}
If $S(l,t) > S(l,t')$ then Attacker attacks target $t$.
\end{lemma}
\begin{IEEEproof}
For a specific control implementation $l$ and two targets $t, 
t'$, \pAw's best response can be found by comparing $e_{lt}, e_{lt'}
$.
If $e_{lt} > e_{lt'} \Leftrightarrow S(l,t) - C(l) > S(l,t') - 
C(l) \Leftrightarrow S(l,t) > S(l,t')$, \pA prefers to attack target 
$t$.
Specifically we define this property as: 
\begin{equation*}\label{deltaSL}
\Delta S(l) := S(l,t') - S(l,t)
\end{equation*}
Therefore, if $S(l,t) > S(l,t') \Leftrightarrow S(l,t') - S(l,t) < 0 
\Leftrightarrow \Delta S(l) < 0$, \pA chooses $t$.
\end{IEEEproof}

\subsection{Saddle Points}
\label{sec_saddle}
Since we are investigating a two-person zero-sum game with finite number of actions for both players, and according to Nash \cite{Nash:NAS:1950} it admits at least a Nash Equilibrium (NE) in mixed strategies.
Saddle-points correspond to Nash equilibria as discussed in \cite{Alpcan:Book:2010}.
The following result, from \cite{Basar:Book:1995}, establishes the existence of a saddle (equilibrium) solution in the games we examine and summarises their properties.

The investigated cybersecurity game admits a saddle point in mixed strategies, $(\Phi^*, \Theta^*)$, with the property 
\[
\Phi^* = \arg \max_{\Phi} \min_{\Theta} J_U(\Phi,\Theta)
\]
\[
\Theta^* =\arg \max_{\Theta} \min_{\Phi} J_A(\Phi,\Theta).
\]

\begin{corollary}\label{cor_maxmin}
Regardless of the Attacker's strategy, the Nash Defender guarantees a minimum performance, that is an upper limit of expected damage.
\end{corollary}
\begin{IEEEproof}
The minimax theorem \cite{minimax} states that for zero sum games NE, maxmin and minimax solutions coincide.
Therefore $\Phi^* = {\tt \arg}\min_{\Phi} \max_{\Theta} J_A (\Phi,\Theta)$.
\end{IEEEproof}

\subsection{Non-zero Sum Games}
\label{sec_non_zero}
Note here that, since in this work we consider zero sum games, two criticisms are possible:

\begin{remark}
\label{rmk_1}
The gain of the Attacker is not, in general, equal to the loss of the defender.
\end{remark}
\begin{remark}
\label{rmk_2}
The Attacker's payoff is not related to the defender indirect costs.
\end{remark}

We address both Remarks by noticing that a significant class of \emph{realistic cybersecurity games} can be mathematically reduced to zero sums games.
Remark 1 is addressed by the following lemma. 

\begin{lemma}
The equilibrium ($\Phi^*,\Theta^*$) in our zero sum cybersecurity game $G$ remains the same in the negative affine transformation of this game in which the Attacker's gain does not equal the Defender's loss.
\end{lemma}
\begin{IEEEproof}
We claim that a model of the \pA where his payoffs are a negative affine transformation of the \pD loss is a reasonable model. 
For example by selling stolen data on the black market for only one tenth of the data's value.

A negative affine transformation of the Defender's $\DefMat$ matrix is defined as $\omega \, \DefMat + \psi$, where $\omega$ is a negative scalar, and $\psi$ is a constant matrix of the same dimension as $\DefMat$.
Therefore, in addition to the cybersecurity game $G=(\DefMat,-\DefMat)$, we intuitively define the negative affinity of this game as $G^-=(\DefMat,\omega \, \DefMat + \psi)$.

Suppose $\Phi^*,\Theta^*$ are the equilibrium strategies in $G$.
First, it is easy to see that $\Phi^*$ is the Defender's equilibrium strategy in both $G$ and $G^-$ due to the Defender's game matrix remaining the same. 
Formally, 
$\Phi \, \DefMat \, \Theta^* \leq \Phi^* \, \DefMat \, \Theta^*$.
Similarly, we prove that $\Theta^*$ is Attacker's equilibrium strategy in both games. 
We have that
$\Phi^* \, (\NegDefMat) \, \Theta \leq \Phi^* \, (\NegDefMat) \, \Theta^*  
\Rightarrow    
\Phi^* \, \DefMat \, \Theta \geq \Phi^* \, \DefMat \, \Theta^* \Rightarrow 
\Phi^* \, (\omega \, \DefMat + \psi) \Theta \leq \Phi^* \, (\omega \, \DefMat + \psi) \Theta^*$. 
This means that equilibria are the same in both $G,G^-$.
\end{IEEEproof}

\begin{lemma}
A game $\widehat{G}$ where the Defender's indirect cost $C$ is a positive affine transformation of the direct cost $S$, has the same maxmin solution with $G$.
\end{lemma}
\begin{IEEEproof}
According to the Lemma we have that in $\widehat{G}$ \pDw's payoff is given by 
\[
S - (\kappa \, S - \mu) = S \, (1-\kappa) - \mu,
\]
where $\kappa, \mu$ are positive scalars.
Assume that at the equilibrium of $\widehat{G}$ \pDw's best response is $\Phi^*$.
Then we have
\begin{align}
&\Phi \, \Big[S \, (1-\kappa) - \mu \Big] \, \Theta^* \leq 
\Phi^* \, \Big[S \, (1-\kappa) - \mu\Big]\, \Theta^* \nonumber\\
&\Rightarrow \Phi \, (S - \kappa \, S - \mu) \, \Theta^* \leq
\Phi^* \, (S - \kappa \, S - \mu) \, \Theta^*  \nonumber \\
&\Rightarrow \Phi \, (S - \mu) \, \Theta^* \leq
\Phi^* \, (S - \mu) \, \Theta^*
\nonumber \\
&\overset{\mu=C}{\Longrightarrow} 
\Phi \, (S - C) \, \Theta^*
\leq
\Phi^* \, (S - C) \, \Theta^*
\Rightarrow
\Phi \, \DefMat \, \Theta^*
\leq
\Phi^* \, \DefMat \, \Theta^*. \nonumber
\end{align}

Therefore $G,\widehat{G}$ have the same equilibria, and from Corollary \ref{cor_maxmin} these are also maxmin solutions.
\end{IEEEproof}

\subsection{A Small Game Example}
\label{sec_toy_game}
\begin{figure}[H]
\vspace{-0.5cm}
   \centering
   \includegraphics[width=3in]{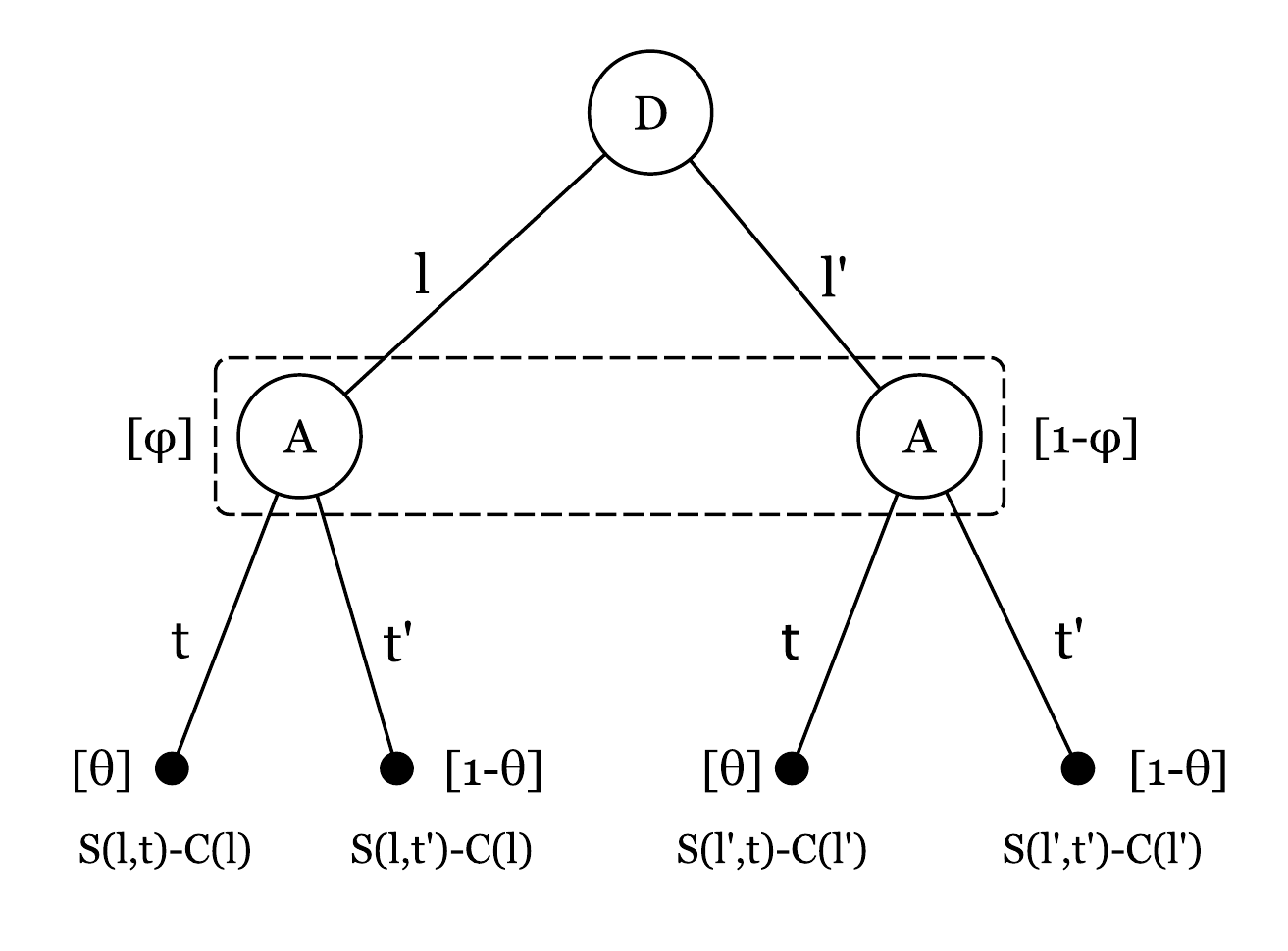}
   \vspace{-0.6cm}
   \caption{Game tree for the control game with 2 implementation levels and two targets.}
   \label{fig:gametree}
\end{figure}

To illustrate the game approach let's consider a toy example consisting of a 2-level, and 2-target control games, where \pD and \pA make their decisions simultaneously, or, equivalently, independently of each other.
The information sets associated with the the control game, investigated in this section, depicted in Fig.\,\ref{fig:gametree}; a dashed curve encircling the \pA nodes has been drawn.
This indicates that \pA cannot distinguish between these two points. In other words, \pA has to arrive at a decision without knowing what \pD has actually chosen. 

\begin{table*}
\centering
\caption{Nash equilibria for the different conditions.}{
  \centering
  \begin{tabular}{ccccc}  
      & \multicolumn{2}{c}{$\DS(t') > \DC$} & \multicolumn{2}{c}{$\DS(t') < \DC$} \vspace{0.1cm} \\
      \hline \vspace{-0.3cm} \\
      \multirow{4}{*}{$\DS(t) > \DC$} & \multirow{2}{*}{$\Delta S(l') > 0$} & \multirow{2}{*}{$(l',t)$} &
      \multirow{2}{*}{$\Delta S(l') < 0$} & \multirow{2}{*}{$(\phi l', (1-\theta) t)$} \vspace{0.3cm} \\  	      
      \cline{2-5} 
          
      & \multirow{2}{*}{$\Delta S(l') < 0$} & \multirow{2}{*}{$(l',t')$} & 
      \multirow{2}{*}{$\Delta S(l) > 0$} & \multirow{2}{*}{$((1-\phi) l, \theta t')$} \vspace{0.3cm} \\
      \hline 
      
      \multirow{4}{*}{$\DS(t) < \DC$} & \multirow{2}{*}{$\Delta S(l) < 0$} & \multirow{2}{*}{$((1-\phi)l', (1-\theta)t')$} & \multirow{2}{*}{$\Delta S(l) > 0$} & \multirow{2}{*}{$(l,t)$} \vspace{0.3cm} \\
      \cline{2-5} 
      
       & \multirow{2}{*}{$\Delta S(l') > 0$} & \multirow{2}{*}{$(\phi l, \theta t)$} & \multirow{2}{*}{$\Delta S(l) < 0$} & \multirow{2}{*}{$(l,t')$} \vspace{0.3cm} \\ \hline
  \end{tabular}}
    \label{tab:best_responses}
\end{table*}

Due to the game being zero-sum, we have kept only the payoffs of \pD at the game tree.
We also defined the mixed strategy of \pD as the probability distribution $[ \phi, 1 - \phi]$, where $\phi$ is the probability of implementing the control at level $l$. 
\pAw's mixed strategy is denoted by $[ \theta, 1 - \theta]$, where \pA chooses to attack $t$ with probability $\theta$.
Table \ref{tab:best_responses} summarizes all possible best responses of the control game for the different conditions discussed in this section.

\subsection{Player Mixed Strategies}
In a two target, two level control sub-game, it is possible to define the probabilities that each player plays a particular mixed strategy.

\begin{lemma}
The Nash equilibrium for a control sub-game for the \pDw's, given by $\phi ^* \in [0,1]$ is:
\begin{equation*}\label{eq:alice_prop}
\phi ^* = \frac{\Delta S(l')}{\Delta S(l') - \Delta S(l)}\end{equation*}
\end{lemma}

\begin{IEEEproof}

The \pD wants to make the \pA indifferent to which target they should 
attack.

This is given by equalising the expected payoff of the \pAw, thus
\begin{eqnarray*}\label{eq:alice_mixPreCon}
  && A(t) = \phi^* \, e_{lt} + (1 - \phi^*) \, e_{l't} \\ \nonumber
  && A(t') = \phi^* \, e_{lt'} + (1 - \phi^*) \, e_{l't'} \nonumber
 \end{eqnarray*}
giving
\begin{equation}\label{eq:alice_mixPre}
  \phi^* \, e_{lt} + (1 - \phi^*) \, e_{l't} = \phi^* \, e_{lt'} + (1 - \phi^*) \, e_{l't'}.
\end{equation}
We can substitute terms such that Eq.\,(\ref{eq:alice_mixPre}) 
can be written in terms of $e_{lt}$, hence
\begin{eqnarray*}\label{eq:eve_sub}
   && e_{l't} = e_{lt}  - \Delta S(t) + \Delta C \\
   && e_{lt'} = e_{lt} - \Delta S(l)\\
   && e_{l't'} = e_{lt} - \Delta S(t) + \Delta C - \Delta S(l') 
\end{eqnarray*}

By substituting the equations above into Eq.\,(\ref{eq:alice_mixPre}) we get
\begin{align}\label{eq:alice_mixSubbedEqual}
  &\phi^* \, e_{lt} + (1 - \phi^*) \, (e_{lt} - \Delta S(t) + \Delta C) =
  \phi^* \, (e_{lt} - \Delta S(l))  \nonumber\\ 
  & + (1 - \phi^*)(e_{lt} - \Delta S(t) + \Delta C - \Delta S(l')).
\end{align}

Eq.\,(\ref{eq:alice_mixSubbedEqual}) can be expanded and reduced to
\begin{equation*}\label{eq:alice_mixSubbedReformed}
  \Delta S(l') = \phi^* (\Delta S(l') - \Delta S(l)).
\end{equation*}

This then gives
\begin{equation*}\label{eq:alice_phi}
  \phi^* = \frac{\Delta S(l')}{\Delta S(l') - \Delta S(l)} 
\end{equation*}
\end{IEEEproof}


\begin{lemma}
The Nash strategy of the \pA in a control sub-game, is given by
\begin{equation*}\label{eq:eve_prop}
\theta ^*= \frac{\Delta S(t) - \Delta C + \Delta S(l') - \Delta S(l)}{\Delta S(l') - \Delta S(l)} 
\end{equation*}
\end{lemma}
\begin{IEEEproof}
At the equilibrium, the \pA wants to make the \pD indifferent to which target they should attack.  
This is given by equalising the expected payoff of the \pDw:
\begin{eqnarray*}\label{eq:eve_mixPreCon}
  && D(l) = \theta^* \, a_{lt} + (1 - \theta^*) \, a_{lt'} \\ \nonumber
  && D(l') = \theta^* \, a_{l't} + (1 - \theta^*) \, a_{l't'}. \nonumber
 \end{eqnarray*}
Therefore
\begin{equation}\label{eq:eve_mixPre}
  \theta^* \, a_{lt} + (1 - \theta^*) \, a_{lt'} = \theta^* \, a_{l't} + (1 - \theta^*) \, a_{l't'}.
\end{equation}

We can substitute terms such that Eq.\,(\ref{eq:eve_mixPre}) can be written in terms of $a_{lt}$, and therefore
\begin{eqnarray*}\label{eq:alice_sub}
   && a_{l't} = a_{lt}  + \Delta S(t) - \Delta C \\
   && a_{lt'} = a_{lt} + \Delta S(l)\\
   && a_{l't'} = a_{lt} + \Delta S(t) - \Delta C + \Delta S(l') 
\end{eqnarray*}
By substituting the equations above into Eq.\,(\ref{eq:eve_mixPre}) we get:
\begin{eqnarray*}\label{eq:eve_mixSubbedEqual}
  &&\theta^* \, a_{lt} + (1 - \theta^*) \, (a_{lt} + \Delta S(l)) = \nonumber\\
  &&\theta^* (a_{lt} + \Delta S(t) - \Delta C) + (1 - \theta^*) \, (a_{lt} + \nonumber \\
  &&\Delta S(t) - \Delta C + \Delta S(l')).
\end{eqnarray*}

The above equation can be expanded and reduced to:
\begin{eqnarray*}\label{eq:eve_mixSubbedReformed}
&&a_{lt} + \Delta S(l) - \theta^* \, \Delta S(l) = a_{lt} + \Delta S(t) - \Delta C +\nonumber \\ 
&&\Delta S(l') - \theta^* \, \Delta S(l'). 
\end{eqnarray*}

This then gives
\begin{equation*}\label{eq:eve_theta}
 \theta^* = \frac{\Delta S(t) - \Delta C + \Delta S(l') - \Delta S(l)}{\Delta S(l') - \Delta S(l)}. 
\end{equation*}
\end{IEEEproof}

\subsection{Numerical Illustration}
\label{sec_numeric_illustration}
We see that \pDw's strategy is derived only from $\Delta S(l)$ and $\Delta S(l')$. This is since \pD wants to make the \pA indifferent to the target they want to attack at the equilibrium.  
In this way the aspects of $\Delta S(t)$ and $\Delta C$ are not represented as they do not impact \pAw.

\begin{figure}[t]
  \begin{center}
  \includegraphics[width=3.7in]{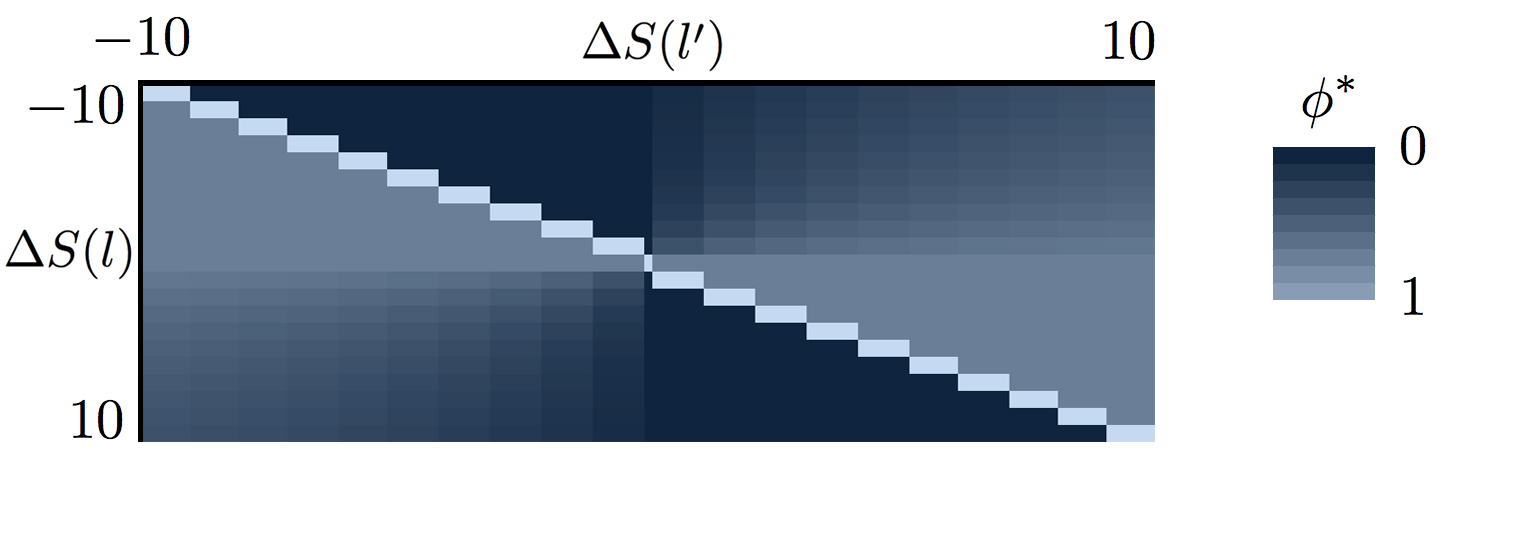}
  \caption{\pDw's Strategy Space}
  \label{fig:phistar}
  \end{center}
\end{figure}

In Fig.\,\ref{fig:phistar} we see the results with the inclusion of the pure strategy solutions. 
When $\Delta S(l) = \Delta S(l')$ the solution is undefined, such that \pD has no incentive to play one defence over the other, as there is no $\phi$ that will influence the outcome of the game.  
Additionally, we see that when $\Delta S(l)$ does not equal $\Delta S(l')$ and both $\Delta S(l) \leq 0$ and $\Delta S(l') \leq 0$  or $\Delta S(l) \geq 0$ and $\Delta S(l') \geq 0$ then we have a collapse onto a single dominant strategy.  The single strategy, represented here by $\phi^* = 0$ or $\phi^* = 1$, is defined depending on which solution would make the attacker the most indifferent to which target they wish to attack.

When $(\Delta S(l) > 0) \, \wedge \, (\Delta S(l') < 0)$ or $(\Delta S(l) < 0) \, \wedge \, (\Delta S(l') > 0)$, then we have the conditions for a mixed strategy.  This is because there is no pure strategy that is dominant, and therefore the defender aims to make the attacker as indifferent as possible.


\section{Cyber Security Investment}
\label{sec:investment}
Thus far the analysis performed has considered a single-control, two-targets, two-levels game. 
Our plan for cyber investment is to solve a set of control sub-games with $n$ targets and up to $m$ control levels for each control. 
Given these game solutions we will then use a Knapsack algorithm to provide the general investment solution.
The control-game solutions provide us with information regarding the way in which each security control is best implemented, so as to maximise the benefit of the control with regard to both the \pAw's strategy, and the indirect costs of the organisation.  

It is easy to see that, in control sub-games, the games look only at the vulnerabilities that are directly relevant to the control being implemented.  
The cyber security investment problem expands to represent all of an organisation's vulnerabilities and selecting the best cyber security controls based on the outcomes of the control-games.

With regard to an implementation of cyber security processes based on the sub-game solutions, it is important to understand what a 
control game solution represents in the process of making those decisions.  
In particular this is about what a mixed strategy means in terms of control implementation.

We motivate the concept of a mixed strategy as a method for trying to define where in the system it is most effective to implement the 
control.  
Based on our interpretation of the structure of a network, this will generally involve protecting devices at the highest level with the strictest controls where possible, then assigning lower levels of controls to devices and users that operate at depths with less sensitive data.

This is performed by creating a logical ordering of the most important devices, based on the perceived risk of the device or the user.  
While there may be a logical ordering across an organisation for all controls, it often might make more sense to order users and devices specifically for each control based on vulnerability.

\begin{example}
 To illustrate this we take for example a security control entitled \emph{Vulnerability Scanning and Automated Patching}, and we assume 5 different implementation levels i.e. $\{0,1,2,3,4\}$ where level 4 corresponds to \emph{real-time scanning} while level 2 to \emph{regular scanning}.
 We say that a mixed strategy $[ 0, 0, 0.7, 0, 0.3 ]$ determines a cyber security plan that dictates the following:
 \begin{itemize}
   \item $0.3 \mapsto$  real-time scanning for the 30\% of the most important devices
   \item $0.7 \mapsto$  regular scanning for the remaining 70\% of devices
 \end{itemize}
 This mixed strategy can be realised more as an advice to a security manager on how to undertake different control implementations rather than a rigorous set of instructions related only to a time factor. 
 We claim that our model is flexible thus allowing the defender to interpret mixed strategies in different ways to satisfy their requirements. 
\end{example}

\subsection{Full Game Representation}
\label{sec_full_game}
A Full Game representation considers the method of solving the investment problem by creating a strategic game containing the set of feasible choices available to both players.  
\pDw's pure strategies are comprised of an implementation level for each of the controls, and \pAw's pure strategies consist of each target in the set of all possible targets.
One of the considerations that needs to be made is with regards to the budget.  
A pure game-theoretic solution for the cyber investment problem would require modelling $n$ targets, $m$ control levels and $c$ controls. 
A naive choice would be to consider $c\times m$ $\times$ $n$ games. 
However it is not clear how to force these game solutions to satisfy budget constraints. 
A game model satisfying budget constraints could be built using the idea of ``schedules'' \cite{Tsai:2009}, i.e. a pure strategy is a tuple of $c\times m$ bits where each bit represents the implementation of a control at a particular level, 1 stands for implemented and 0 for not implemented. 
The budget requirement can be easily imposed on such tuples, for example by only considering tuples whose costs do not exceed the budget. 
The problem with this proposal is that, in principle, there could be an exponential number of pure strategies, in the order of $2^{(c\times m)}$. 
Also it would be non-trivial to choose appropriate payoffs for such tuples.
In this case, we restrict the combination of controls in the payoff matrix to only those that can be purchased based on the maximum amount of budget.

\subsection{Pure Knapsack Representation}
\label{sec_pure_knap}

A Pure Knapsack representation considers the method of solving the investment problem given that \pD only considers the implementation of ``whole'' controls.  We have chosen a 0-1 Multiple Choice, Multi-Objective Knapsack.

The choice of this type of Knapsack is motivated as follows: ``0-1'' because each level of implementation of a control is implemented or not implemented, ``Multiple Choice'' because only one solution for each control (the optimal one) ought to be chosen and ``Multi-Objective'' because each target represents an optimisation objective.
More precisely we define in this case as $Q$ a control implementation, where $Q_{jl}$ is the implementation for control $j$ at level $l$.

We define a solution to the Knapsack problem as
\begin{equation*}\label{eq:knap_solution}
 \Psi = \{Q_{1l_1} \dots Q_{cl_c} \}, l_i \in \levels.
\end{equation*}

A solution $\Psi$ takes exactly one level for each control as a policy for implementation, where notice for each control there exists a solution $Q_{j0}$, which dictates that control $j$ should not be used. 

To represent the cyber security investment problem, we need to expand the definitions for both expected damage ($S$) and effectiveness ($E$) to incorporate multiple controls.  So we expand $S$ such that $S(\Psi, t)$, which is the expected damage on target $t$ given the implementation of the levels associated with $\Psi$.  
Likewise, we expand the definition of the effectiveness of the implemented solution on a given target as $E(\Psi, t)$.  
Additionally, we consider $\Gamma(Q_{jl})$ as the direct cost of implementing $Q_{jl}$, and the maximum of all $\Gamma$s cannot exceed the available cyber security budget $B$.
Then we solve 
\begin{eqnarray}\label{eq:investment}
&\max\limits_{\Psi}\min\limits_{t_i} & \{1 - \sum_{j=1}^{c}\sum_{l=0}^{m} E(Q_{jl}, t_i) z_{jl}\} \, S(\Psi, t_i) - C_{j}(l) \nonumber \\
&\text{s.t.}& \, \sum_{j=1}^{c} \sum_{l=0}^{m}\Gamma(Q_{jl}) \, z_{jl} \leq B\nonumber\\
&&\sum_{l=0}^{m} z_{jl} = 1, \, z_{jl} \in \{0,1\}, \, \forall j = 1, \dots, c \nonumber
\end{eqnarray}
where $z_{jl}$ is the probability of implementing control $j$ at level $l$, and $B$ is the available cyber security budget.

\subsection{Hybrid Method}
\label{sec_hybrid}
The Hybrid approach avoids the problems of the Full Game method by considering the particular game solutions for each control as part of an overall combinatorial optimisation problem which we also solve using a 0-1 Multiple Choice, Multi-Objective Knapsack.

In this case we define as $Q$ a sub-game solution, where $Q_{jl}$ is the sub-game solution for control $j$ implemented at level $l$.

We represent the 0-1 Multiple Choice, Multi-Objective Knapsack Problem as:
\begin{eqnarray}\label{eq:investment}
&\max\limits_{\Psi}\min\limits_{t_i}&\{1 - \sum_{j=1}^{c}\sum_{l=0}^{m} E(Q_{jl}, t_i) z_{jl}\}\,S(\Psi, t_i) \nonumber \\ 
&\text{s.t.}&\,\sum_{j=1}^{c} \sum_{l=0}^{m}\Gamma(Q_{jl}) \, z_{jl} \leq B \nonumber\\
&&\sum_{l=0}^{m} z_{jl} = 1, \, z_{jl} \in \{0,1\},\,\forall j = 1, \dots, c \nonumber
\end{eqnarray}

Notice that in the above formula, we do not have the factor of indirect cost ($C_{j}(l)$), this is because indirect cost is taken into consideration in the control sub games.

In addition, we consider a tie-break condition in which if multiple solutions are viable, in terms of maximising the minimum, according to the above function we will select the solution with the lowest cost.  
This ensures that an organisation is not advised to spend more on security controls than would produce the same net effect.

\section{Case study:\,SANS Critical Security Controls, CWE Top Software Vulnerabilities}
\label{sec_case}
To compare the Full Game, Hybrid and Pure Knapsack methods of decision support, we have developed a sample case study similar to one expected in a real environment.  

In this work, we are interested in comparing two aspects of the solutions generated by the different methods.
The first is the optimality of the solutions, and the other is their complexity.
To this end, we consider the \emph{optimality of the solution} to be the expected damage of the implemented set of controls at the weakest level, for a given budget.

The \emph{complexity of the solution} provides a pivotal role in decision support with cyber security, where overly complex solutions are potentially difficult to implement and follow.  
This is relevant with mixed strategy equilibria.

\subsection{Modelling SANS Critical Controls and CWE Top Software Vulnerabilities}
Our case study is created using a mapping from the SANS Critical Security Controls \cite{web:SANS} combined with the CWE Top 25 Software Vulnerabilities\cite{web:CWE}.  
In this mapping, we define a control as a collection of any of the associated processes defined by a single critical security control in the SANS top 20.  Additionally, we consider a vulnerability as any of the software vulnerabilities that are defined in the CWE Top 25.  
Using data associated with these two sources we are able to build the core components of a case study to test our methodology.

From the SANS Critical Security controls, we define a level of Implementation for a given control as a single action point listed for the control.  
For each control, the control levels are considered in order, based on their position in the list.  
In some cases where there is significant overlap between control levels, levels can be combined.
This is aimed at reducing the number of strategies and computational complexity of the problem.

Using the classifications provided by the CWE Top 25 Software vulnerabilities we are able to categorise the different classes of vulnerabilities that each of the controls is able to mitigate.  
CWE proposes three categories \emph{Insecure Interactions Between Components}, \emph{Risky Resource Management} and \emph{Porous Defences}.  
A given vulnerability falls into one of the three categories and we consider that any control may cover the vulnerabilities associated with one or more of the categories.

To calculate part of the risk, we consider the threat value to be directly associated with each of the vulnerabilities. 
CWE defines a score out of 100 for a number of vulnerabilities, with those scoring highest published in their top 25.  
The damage values associated with each of the weaknesses have been scaled to fit within the range of the other values gathered for this case study.

The efficiency of a control level is considered to be a reduction in the effectiveness of a given control based on the idea that a control should be effective at stopping an attack.  
In reality an \pA may be able to circumvent these controls.
We consider that their ability to do so is linked to the availability and ease with which information about vulnerabilities can be discovered.
Then, we directly link this ability to the attack factors provided by CWE.  
To do this, we apply a weighted percentage of the values for the four key factors that CWE defines regarding a vulnerability capped according to the level.  
The four attack factors that CWE defines for a given vulnerability are \emph{Prevalence}, \emph{Attack Frequency}, \emph{Ease of Detection} and \emph{Attacker Awareness}.

CWE provides an expected cost value to repair the vulnerability for each weakness.  
For this mapping, we consider each cost separately for each vulnerability.
More specifically, the direct cost of a control is given as the sum of all the costs for the vulnerabilities that it covers. 
This direct cost value is considered to be the cost of implementation at the highest level for the control.  
The direct costs for lower levels are scaled uniformly based on the number of levels the control has.

The mapping provided is able to cover the technical aspects of the controls and vulnerabilities, however, there are certain aspects unique to each organisation .  
We consider that both the impact of a successful attack and the indirect costs incurred need to be defined based on priorities and requirements of the organisation.  
The impact of a successful attack is given by not only the data loss to the organisation, but also by the loss of reputation, as well as other costs associated with a breach.  
The indirect costs are considered to be the importance that the organisation places on the day to day performance of the system, as well as the ability and willingness of staff to comply with any additional security policies.

\subsection{Values}
The case study presented in this work considers a network separated into three different depths, consistent with Fig.\,\ref{fig:environment}, where \pD has seven different controls available to protect the network from twelve different vulnerabilities.
For this example, we consider the levels available to \pD to consist of the quick win processes provided by SANS.
The seven controls are shown in Table \ref{tab:DControls} and the twelve vulnerabilities are shown in Table \ref{tab:vuln_factors}.
Based on the controls used, the budget at which all controls become available at the highest level equals 82.

\begin{table*}
\centering
  \renewcommand{\arraystretch}{1}
  \caption{Case Study Controls.\label{tab:DControls}}{
  \begin{tabular}{|c|c|}
   \hline
   \textbf{Control} & \textbf{Levels}  \\  \hline
   Inventory of Authorised and Unauthorised Devices (1) & 3 \\
   Inventory of Authorised and Unauthorised Software (2) & 3 \\
   Secure Configuration for Hardware and Software on Devices (3) & 5 \\
   Continuous Vulnerability Assessment and Remediation (4) & 4 \\
   Malware Defences (5) & 6 \\
   Application Software Security (6) & 2 \\
   Controlled Use of Administrative Privileges (12) & 6 \\
\hline
\end{tabular}}
\end{table*}

\begin{table*}
	\renewcommand{\arraystretch}{2}
	\centering
	\caption{Case Study Vulnerabilities.\label{tab:vuln_factors}}{
	\begin{tabular}{|l|c|c|c|c|l|c|c|c|c|}
		\hline
		\parbox[l]{\boxsize}{$v_z$:\,\textbf{Vulnerability} \par \textbf{(CWE-code)}}
		& \textsf{PR} 		
		& \textsf{AF} 	
		& \textsf{ED} 	
		& \textsf{AA}	
		&\textbf{Vulnerability} 	
		& \textsf{PR} 		
		& \textsf{AF} 	
		& \textsf{ED} 	
		& \textsf{AA}	
		\\\hline
		$v_1$:\,\textsf{SQLi} (89)
		&2 	&3	&3	&3 & 
		\parbox[l]{\boxsize}{$v_7$:\,$\tt Missing$ \par $\tt encryption$ (311)}
		&2	&2	&3	&2		
		\\\hline
		\parbox[l]{\boxsize}{$v_2$:\,$\tt OS\,command$ \par $\tt injection$ (78)}
		&1	&3	&3	&3 &
		\parbox[l]{\boxsize}{$v_8$:\,$\tt Unrestricted$ \par $\tt upload$ (434)}
		&1	&2	&2	&3
		\\\hline
		\parbox[l]{\boxsize}{$v_3$:\,$\tt Buffer$ \par $\tt overflow$ (120)}
		&2	&3	&3	&3 &
		\parbox[l]{\boxsize}{$v_9$:\,$\tt Unnecessary$ \par $\tt privileges$ (250)}
		&1	&2	&2	&2
		\\\hline
		$v_4$:\,\textsf{XSS} (79)
		&2	&3	&3	&3
		&$v_{10}$:\,\textsf{CSRF} (352)
		&2	&3	&2	&3
		\\\hline
		\parbox[l]{\boxsize}{$v_5$:\,$\tt Missing$ \par $\tt authentication$ (306)}
		&1	&2	&2	&3 &
		\parbox[l]{\boxsize}{$v_{11}$:\,$\tt Path$ \par $\tt traversal$ (22)}
		&3	&3	&3	&1
		\\\hline
		\parbox[l]{\boxsize}{$v_6$:\,$\tt Missing$ \par $\tt authorization$ (862)}
		&2	&3	&2	&2 &
		\parbox[l]{\boxsize}{$v_{12}$:\,$\tt Unchecked$ \par $\tt code$ (494)}
		&1	&1	&2	&3		
		\\\hline
	\end{tabular}}
\end{table*}

\subsection{Optimality Comparison}
\begin{figure}[H]
  \includegraphics[width=3.3in]{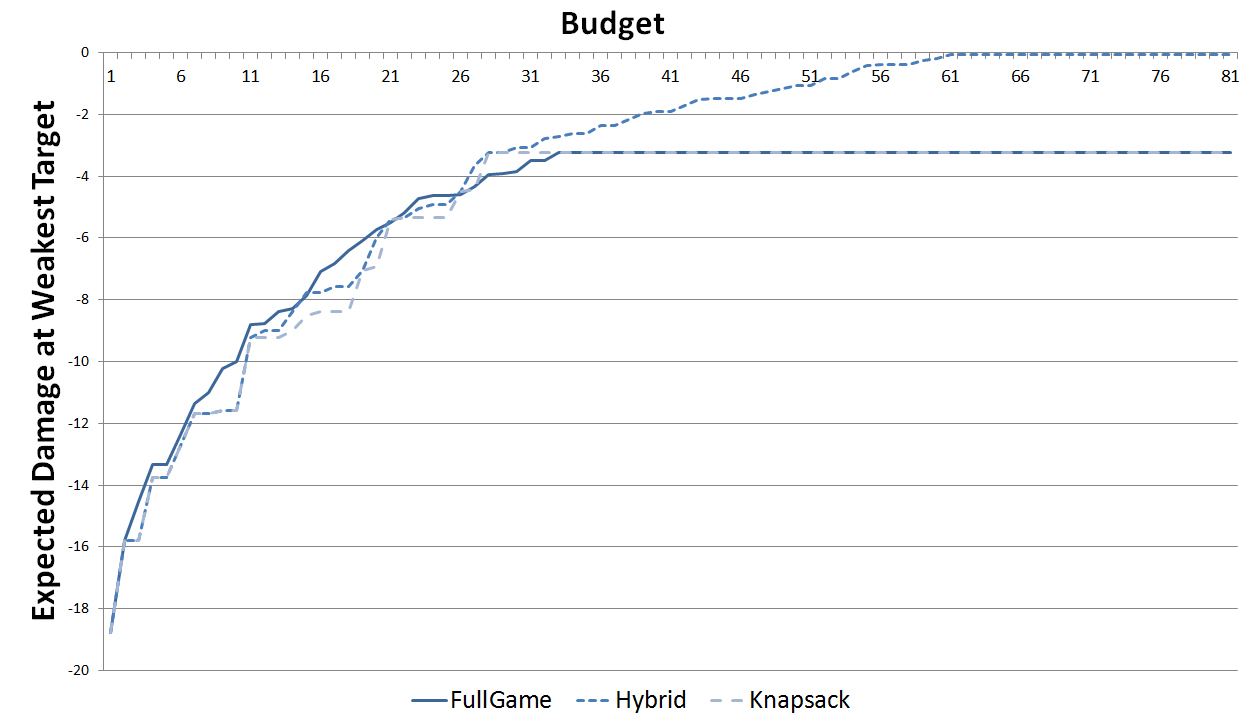}
  \caption{Case Study Results with Normally Expected Indirect Costs}
  \label{fig:results_normal}
\end{figure}

In comparing the damage at the weakest target provided by the Full Game, Hybrid Method to the Knapsack Representation, we can see in Fig.\,\ref{fig:results_normal} that, in general, the Full Game Representation will provide a better defence to the weakest target for low budget levels.  
However, once the budget becomes larger, we see that the Hybrid Method is able to reach a level of coverage that will minimise the damage at each target, whereas neither the Full Game Representation nor the pure Knapsack Representation fully cover the weakest target, even with the maximum budget.  
This is owing to the impact that the indirect cost has on the decision-making process.  
Where the Hybrid Method includes the impact of direct cost in the decisions, regarding the optimality of the deployment of the control at each level
\begin{itemize} 
  \item the pure Knapsack includes the indirect cost, as a whole, in the outcome of the optimisation, and
  \item the Full Game applies the indirect cost to each strategy in the payoff matrix.
\end{itemize}
 
We can also see that the impact of the indirect cost causes the Full Game Representation to become inefficient compared to the Hybrid and Knapsack, before reaching the maximum defence.  
This occurs in Fig.\,\ref{fig:results_normal} within the budget range of 27 - 32.

Fig.\,\ref{fig:results_normal} shows that with low indirect costs, the outcome of the control sub-games allow for the availability of better strategies with lower budgets than the Knapsack-only representation.  
This is due to the Hybrid Method being able to use a combination of packages that has a lower direct cost, but it provides the required coverage to the weakest target. 
Where a similar coverage of the weakest target is only available to the Knapsack when the pure strategies are made available.

It has also been seen that with higher indirect costs both the Full Game and pure Knapsack Representation will offer increasingly poor results when compared to the Hybrid Method.  
This is due to the point at which the cost of implementing controls outweighing the benefit being reached at a lower budget. 

\begin{figure}[h]
  \includegraphics[width=3.3in]{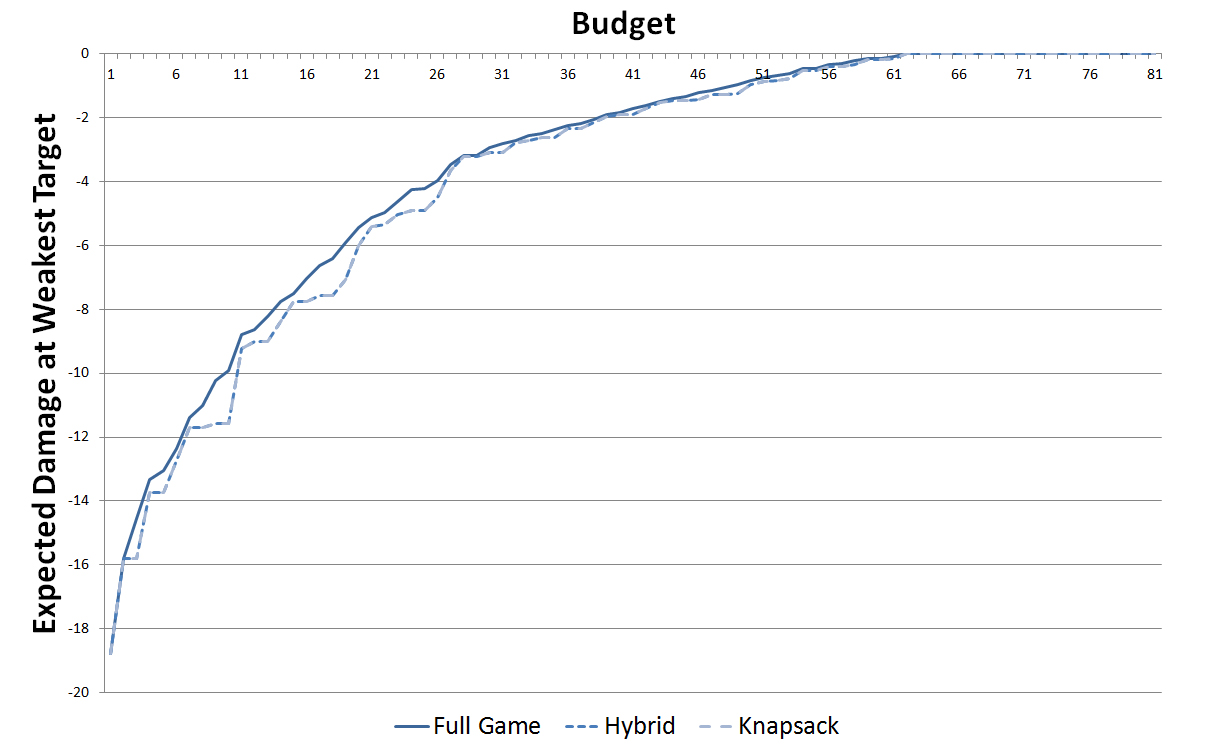}
  \caption{Case Study Results with No Indirect Costs}
  \label{fig:results_none}
\end{figure}

In Fig.\,\ref{fig:results_none}, we see that when there are no indirect costs, the Hybrid Method and Knapsack Representation-only method have exactly the same output.  
This is due to the outcome of each control sub-game providing a pure strategy at the highest level of implementation, which would result in the Knapsack solver having identical strategies with regard to the optimisation.

With regard to the indirect cost, if there is no indirect cost to implementing the control then there is no trade-off to a higher defence.  
This means that providing that an appropriate budget is available, then the best defence will be purchased by all methods, as seen in Fig.\,\ref{fig:results_none}.  
Most importantly, this means that the Full Game Representation provides solutions that are often more optimal, but at least no worse than those generated by the other methods.  
This is because the Full Game Representation has no drawbacks to the implementation of the best controls in the most optimal configuration, which is still a restriction on the two methods that implement the 0-1 restriction of the Knapsack.

\subsection{Complexity Comparison}
To identify how complex a solution is, we need to consider the composition of solutions to assess how complex the solutions are and if they advice could be reasonably followed by an individual on behalf of an organisation.

If we take for example a budget of 18, the solution provided by the Full Game Representation, comprises of a mixed strategy consisting of 4 packages.  
If we consider the three major strategies in Table \ref{tab:18Solution}, then we can see that they all suggest the use of control 5 at level 2.  
Additionally, we see that at a minimum, control 6 should be implemented at level 1, with an increase to level 2 for 23\% of the time.
This suggests always having software versions supported by vendors, but only implementing web application firewalls on the top 23\% of the system.

We can also see that the solution suggests using control 2 and control 7 with 41.3\% and 76.8\% respectively. 
Both of these controls determine restricting access, either through application whitelisting or a reduction in non-essential admin rights and activities.  
Control implementations can be considered in one of two ways, either the percentage relates to the number of devices that feature this control or the severity with which the control is implemented.  
In the case of application whitelisting an implementation level of 41.3\% would allow a greater degree of software availability to users than a higher percentage.

Control 3 suggests using level 2 35.5\% of the time, with level 1 not being suggested individually.  
This relates to using both secure configurations for each operating system and implementing automated patching tools.  
The easiest way to interpret this solution is to state that the top 35.5\% of devices utilise both levels of this control, while all other devices don't.

While in this case the mixed strategies provided by the Full Game Representation do not represent vastly different strategies, the addition of more controls and vulnerabilities will increase the complexity of the solution space. 

\begin{table}[t]
\centering
  \renewcommand{\arraystretch}{1.3}
  \caption{Full Game Solution for a Budget of 18.\label{tab:18Solution}}
  {
  \begin{tabular}{|c|c|}
   \hline
   \textbf{Package} & \textbf{Probability} \\
	$[0,0,0,0,2,2,0]$ & 0.23 \\
	$[0,0,2,0,2,1,1]$ & 0.355 \\
	$[0,1,0,0,2,1,1]$ & 0.413 \\
	$[0,1,1,0,0,1,1]$ & 0.001 \\
    \hline
\end{tabular}
}
\end{table}

Additionally for a budget of 18, we found that the Hybrid Method suggests using the solution $[0,1,1,0,4,0,1]$, while the pure Knapsack solution suggests $[0,0,2,0,2,1,0]$.  
In this example the package suggested by the Hybrid Method is viable as a solution for Knapsack, as none of the sub-game solutions differ from the highest level pure strategy that would be available to the \pDw.  
This further highlights that the indirect costs are pivotal in demonstrating the optimality of these results.

With a budget of 29, we see that both the Hybrid and the Pure Knapsack select the package $[0,1,1,0,2,2,1]$.  
While the Full Game Representation is able to use this package, it selects it with probability $p = 0.001$.  
This again shows the importance that indirect costs can have on the optimality of the solution, given that while feasible, the game-theoretic method considers it too costly to implement.

If we consider the case when the budget is 48, the Hybrid Method provides the solution $[0,2,5,0,4,2,2]$, where for control 3 (Secure Configuration for Hardware and Software on Devices) the outcome is to use a mixed strategy.  
The mixed strategy suggests using level 4 with $p = 0.609$ and level 5 with $p = 0.391$. 
At level 4 we consider the following of strict configuration management for creating secure images of each OS, and level 5 is concerned with the storage of these images.  In this case we can consider that at all times the secure image is used, but the secure storage of master images is only considered for approximately 40\% of the time.

The pure Knapsack has solutions that can be followed intuitively as they only ever consider a single level of implementation.  
We can also see that the Hybrid Method often uses pure strategies as in many cases the outcomes of the control sub-games lead to a single strategy at many levels.  However, we find that there is an additional level of complexity in the comprehension of the strategies that are produced by the Full Game.  
Such complexity can potentially lead to strategies that can not easily be followed by a user to gain the most from the solution.  
In these cases, there is a risk that the solutions are not followed correctly and with security.
This could lead to a potentially weaker defence over a seemingly weaker, but more easily interpreted solution.
\section{Conclusions}
In this paper we have presented an analysis of a hybrid game-theoretic and optimisation approach to the allocation of an organisation's cyber security budget.
For this purpose, we have compared three different approaches to allocating this budget.
We found that when there are no indirect costs to consider or the indirect costs have a minimal impact compared to the benefit, then the Full Game Representation gave solutions better than or equal to those of the other two methods. 
However, when an increase in security is matched by the indirect costs, then the Full Game Representation, is not able to overcome the addition of the indirect costs in favour of a stronger defence, in a similar way to the Pure Knapsack Representation.  The Hybrid Method, however, considers the indirect costs as part of each control game and therefore considers the optimality of each control first, and that an optimal solution is the best valid combination of the optimised controls.

In terms of understanding the solutions, we have found that with a relatively small case study the results can be interpreted in a relatively simple manner.  
However, we are concerned that for a larger case study the Full Game Representation would create solutions that are too complex to be interpreted in an accurate manner so that they could result in a weaker defence. 

This work also highlights the impact, which the indirect costs have on the problem of cyber security budget allocation.  
Considering the downside that the implementation of a control may have on the organisation is important, since it can better capture the decision-making process required for investment.  The results presented in this paper demonstrate how those indirect costs are able to influence the optimal decision in cyber security budget allocation.

We aim to use the work presented in this paper to inform models of attacks against a system.  
These games model the interactions between an attacker and defender at the \emph{Point of Attack}, and during an ongoing attack.  
To do this we will consider multi-stage games which represent the stages of an attack and recovery in a system.  
The techniques presented in this paper should allow for the development of tools for better allocation of resources to help prevent successful attacks made against a system.

In addition, we aim to work with security practitioners in order to create a more detailed case study, and to highlight the operation of this method in a realistic setting.  
The objective of this is to better understand the parameters that organisations have, and how these can best be applied within our framework.  
The result of this analysis will lead to the development of a dedicated tool for cyber security budget allocation.

\bibliographystyle{IEEEtran}
\bibliography{references}
\end{document}